\documentclass[aps,prb,onecolumn]{revtex4}
\usepackage{graphicx}
\usepackage{amssymb}
\usepackage{amsfonts}
\usepackage{amsmath}
\usepackage{bm}
\usepackage{color}
\usepackage{pstricks}
\usepackage{pst-node}

\DeclareMathOperator{\arctanh}{arctanh}

\renewcommand{\vec}[1]{\mathbf{#1}}

\newcommand{\bis}{{\prime\hspace{-0.1ex}\prime}}
\newcommand{\msms}{\scriptscriptstyle{--}}
\newcommand{\psms}{\scriptscriptstyle{+-}}
\newcommand{\msps}{\scriptscriptstyle{-+}}
\newcommand{\psps}{\scriptscriptstyle{++}}

\newcommand{\Xiso}{{\ensuremath{\Xi_{\text{so}}}}}

\newcommand{\pd}{\ensuremath{p\text{-}d}}
\newcommand{\sd}{\ensuremath{s\text{-}d}}
\newcommand{\Pex}{\ensuremath{P_\text{ex}}}
\newcommand{\kp}{\ensuremath{k\cdot p}}

\newcommand{\spu}{|\hspace{-0.25em}\uparrow\rangle}
\newcommand{\spd}{|\hspace{-0.25em}\downarrow\rangle}

\begin{document}

\title{L\"owdin calculus for multiband Hamiltonians}

\author{Agnieszka Werpachowska} \email{rilwen@gmail.com} \affiliation{London, United Kingdom}

\begin{abstract}
This appendix to the paper by Werpachowska and Dietl~\textit{Theory of spin waves in ferromagnetic (Ga,Mn)As}\cite{Werpachowska:2010_PRBa} is a response to inquiries about the derivation of the L\"owdin calculus and its numerical implementation. It presents the detailed version of the L\"owdin calculus for the multiband Hamiltonian taking into account both the $\pd$ and $\sd$ exchange couplings. This should explain how to implement the $s\pd$ exchange interaction in the tight-binding computational scheme. I also include the complete procedure of the Bogoliubov transform for systems with spatial inversion asymmetry.
\end{abstract}
\maketitle

\section{Effective Hamiltonian}
\label{sec:selfconloewdin}

We consider the ferromagnetic phase of a system consisting of $P$ carriers and $N$ magnetic lattice ions, described by the Hamiltonian $\mathcal H_0$ and coupled by the $s\pd$ exchange interaction $\mathcal H'$,
\begin{equation}
\begin{split}
\label{eq:app_Hamiltonian}
\mathcal H = \mathcal{H}_0 + \mathcal{H'} = \mathcal H_0 + \sum_{i=1}^P \sum_{j=1}^{N}  \, \vec{s}_i \cdot \vec{S}_j I(\vec{r}_i - \vec{R}_j) \ ,
\end{split}
\end{equation}
where $\vec{s}_i$ and $\vec{S}_j$ are the $i$-th carrier's and $j$-th ion's spin operators, while $\vec{r}_i$ and $\vec{R}_j$ are their respective positions. The matrix elements of $I$ between the $s$- and $p$-type basis functions ($\alpha$ and $\beta$, respectively) determine the strength of the $\pd$ and $\sd$ exchange interaction, respectively. They are defined as
\[
\alpha = \int_V \mathrm{d}^3r S(\vec{r})^\ast I(\vec{r} - \vec{R}) S(\vec{r}) \ ,
\]
and
\begin{equation}
\label{eq:betadef}
\beta = \int_V \mathrm{d}^3r X(\vec{r})^\ast I(\vec{r} - \vec{R}) X(\vec{r}) = \int_V \mathrm{d}^3r Y(\vec{r})^\ast I(\vec{r} - \vec{R}) Y(\vec{r}) = \int_V \mathrm{d}^3r Z(\vec{r})^\ast I(\vec{r} - \vec{R}) Z(\vec{r}) \ ,
\end{equation}
where $S(\vec{r}),X(\vec{r}),Y(\vec{r})$ and~$Z(\vec{r})$ are the periodic parts of the $s$- an $p$-type band wavefunctions at $\vec{k}=0$. Their values for (Ga,Mn)As are $N_0 \beta = -1.2$~eV and $N_0 \alpha = 0.2$~eV, where $N_0$ is the concentration of cation sites~\cite{Dietl:2001_PRB,Oszwaldowski:2006_PRB,Strahberger:2000_PRB}. In the absence of external fields, $\mathcal H_0$ depends on the carriers' degrees of freedom only.

The different signs of $\pd$ and $\sd$ exchange integrals lead to antiferromagnetic coupling of the ion spins with the valence band $p$-type holes and conduction band $s$-type electrons, respectively. Within the mean-field and virtual crystal approximations, these couplings produce a Zeeman-like spin splitting of the energy bands, described by the additional term in the one-particle Hamiltonian
\[
h = h_0 + x N_0 S ( \beta P_p + \alpha P_s ) s^z \ ,
\]
where $h_0$ describes the host band structure, $x$ is the fractional concentration of Mn ions with total spin $S=5/2$ and $P_p$, $P_s$ are projection operators associated with the $p$- or $s$-type band wavefunctions at $\vec{k}=0$, respectively, which commute with the spin operator. As the $\sd$ interaction strength $\alpha$ is much smaller than $\beta$, the latter determines the strength of the ion-carrier coupling. It is convenient to write $h$ as
\begin{equation}
\label{eq:app_H1part}
h = h_0 + \Delta s^z \Pex \ ,
\end{equation}
where $\Delta$ is the exchange spin splitting of the heavy hole $p$-type bands in the $\Gamma$ point, equal -0.15~eV for $x = 5$\% and rescaled linearly for different Mn concentrations, and $\Pex = P_p + \frac{\alpha}{\beta} P_s$ commutes with the spin operators.\footnote{In general, $\Pex$ is not a projection operator, as $(P_p + \frac{\alpha}{\beta} P_s)^2 = P_p^2 + \frac{\alpha}{\beta}(P_p P_s + P_s P_p) + \frac{\alpha^2}{\beta^2} P_s^2 = P_p + \frac{\alpha^2}{\beta^2} P_s \neq \Pex$ for $\alpha \neq \beta$. However, it maps its image onto itself.} With the above definition of $\Pex$, $h$ is directly applicable to the  \kp\ band structure calculation methods we have used in Ref.~\onlinecite{Werpachowska:2010_PRBa}, as their basis functions are exactly the ones that operators $P_p$ and $P_s$ project on. However, the tight-binding approach uses localised, atomic-like orbitals wavefunctions as a basis. The Mn ions are coupled to $p$-type orbitals of As and $s$-type orbitals of Ga. The spin splittings applied to these orbitals must take into account their weights in the $s$- and $p$-type band wavefunctions close to the centre of the Brillouin zone. The operator $\Pex$ is therefore redefined to
\[
\Pex = \Delta_{5\%}^{-1} \left( e_p P_p^{As} + e_s P_s^{Ga} \right) \ ,
\]
where $\Delta_{5\%} = -0.15$~eV is the heavy hold band spin splitting for 5\% Mn concentration, and $P_p^{As}$ and $P_s^{Ga}$ are the projection operators on the respective orbitals, and the effective orbital spin-splittings for 5\% Mn concentration are
\begin{center}
\begin{tabular}{ccc}
\hline
Parametrisation & $e_p$ [eV] & $e_s$ [eV] \\
\hline
Jancu TBA & -0.2764 & 0.0548 \\
Di Carlo TBA & -0.1940 & 0.0356 \\
\hline
\end{tabular}
\end{center}
The values of the above constants were chosen so that the heavy-hole splitting $\Delta$ matches the one resulting from the \kp\ method (see Fig.~1 in Ref.~\onlinecite{Werpachowska:2010_PRB}). The weights of the projection operators are almost the same in both methods,
\[
\frac{e_{p,\text{ Jancu}}}{e_{s,\text{ Jancu}}} \approx \frac{e_{p,\text{ Di Carlo}}}{e_{s,\text{ Di Carlo}}} \approx \frac{\beta}{\alpha} \ .
\]

The dynamics of magnetic ions coupled to the system of carriers requires a self-consistent description, which takes into account how the carriers react to the ions' magnetization changes. Therefore, we use the L\"{o}wdin perturbation method specifically adapted for multiparticle Hamiltonians,\cite{Loewdin:1951_JCP, Thijssen:2007_B, Ziener:2004_PRB, Werpachowska:2006_MS} to derive an effective Hamiltonian $\mathcal{H}^\text{eff}$ for ions only. The perturbation in question is the $s\pd$ interaction between carriers and ions.

We choose the multiparticle basis states of $\mathcal H$ as $M \otimes \Gamma$. The ion part $M$ has spins quantized along the $z$ direction, while the carrier part $\Gamma$ is a Slater determinant of $P$ one-particle eigenstates $\psi_{\vec{k},m}$ of the one-particle carrier Hamiltonian~\eqref{eq:app_H1part}. The subscripts $\vec{k}$ and $m$ denote the wave vector and the band number, respectively.

The L\"{o}wdin calculus consists in dividing the multiparticle basis states into two subsets, $A$ and $B$,
\[
\mathcal{H} = \begin{pmatrix}
\mathcal H_{AA} & \mathcal H_{AB} \\
\mathcal H_{BA} & \mathcal H_{BB}
\end{pmatrix} \ .
\]
Set $A$ contains all states $M \otimes \Gamma_0$, where $\Gamma_0$ is a Slater determinant of the $P$ lowest eigenstates of $h$. Set $B$ contains all the remaining states, in which at least one carrier is excited above the Fermi level. We construct the effective Hamiltonian for the states from set $A$ only, adding their coupling with set $B$ as a second order perturbation,
\begin{equation}
\label{eq:app_Heff}
\mathcal{H}^\text{eff}_{nn'} = (\mathcal{H}_0)_{nn'} + \mathcal H_{nn'}' + \sum_{n'' \in B} \frac{\mathcal H_{nn''} \mathcal H_{n''n'}}{E - \mathcal{H}_{n''n''}} \ ,
\end{equation}
where $\mathcal{H}_{nn'} = \langle n | \mathcal{H} | n' \rangle$ (similarly for $(\mathcal{H}_0)_{nn'}$ and $\mathcal H_{nn'}'$) and $n, n' \in A$. The term $(\mathcal{H}_0)_{nn'}$ is independent of the ion configurations, so we set it to zero for simplicity. Thus, the effective Hamiltonian $\mathcal{H}^\text{eff}$ depends only on the ion degrees of freedom but, thanks to the L\"{o}wdin method, takes carrier excitations into account and can be used to calculate the spin-wave dispersion in a self-consistent manner.

The variational part of the presented method consists in searching for the energy $E$ somewhere in the region for which we want accurate results, which in our case are the lowest eigenenergies of $\mathcal H$, in particular the groundstate. For a known average spin splitting $\Delta$, we can set $E$ to the total energy of the carrier multiparticle state $\Gamma_0$, $E_{\Gamma_0} = \sum_{(\vec{k}, m) \in \Gamma_0} E_{\vec{k} m}$, where $E_{\vec{k} m}$ is the eigenenergy of $\psi_{\vec{k} m}$, and the sum goes over all occupied eigenstates in $\Gamma_0$. The states $n''$ are of the form $M'' \otimes \Gamma''$, $\Gamma'' \neq \Gamma_0$. To simplify the sum over $n''$, we approximate the diagonal matrix element $\mathcal{H}_{n''n''}$, which depends on both $M''$ and $\Gamma''$, by the total energy of the multiparticle carrier state $\Gamma''$, $E_{\Gamma''} = \sum_{(\vec{k}, m) \in \Gamma''} E_{\vec{k} m}$. It describes the interaction of $\Gamma''$ with the average configuration of the ions' spins corresponding to the spin splitting $\Delta$. We can thus write the Hamiltonian~\eqref{eq:app_Heff} in the following form:
\begin{equation}
\label{eq:app_Heff2}
\mathcal{H}^\text{eff}_{nn'} = \mathcal H_{nn'}' + \sum_{M''} \sum_{\Gamma'' \neq \Gamma_0} \frac{\mathcal H_{nn''} \mathcal H_{n''n'}}{ E_{\Gamma_0} - E_{\Gamma''} } \ , \ n, n' \in A\ .
\end{equation}
The factor $\mathcal H_{nn''} \mathcal H_{n''n'}$ under the sum can be written as $\langle M \otimes \Gamma_0 | \mathcal{H} | \Gamma'' \otimes M'' \rangle \langle M'' \otimes \Gamma'' | \mathcal{H} | \Gamma_0 \otimes M' \rangle$, where $n = M \otimes \Gamma_0$ and $n' = M' \otimes \Gamma_0$. Since the denominator in Eq.~\eqref{eq:app_Heff2} is independent of $M''$, summing over $M''$ is equivalent to inserting an identity operator in the ion Hilbert space, which allows us to write the last term as
\[
\sum_{\Gamma'' \neq \Gamma_0} \frac{ \langle M \otimes \Gamma_0 | \mathcal{H} | \Gamma''  \rangle \langle \Gamma'' | \mathcal{H} | \Gamma_0 \otimes M' \rangle }{ E_{\Gamma_0} - E_{\Gamma''} } \ .
\]
We can thus treat $\mathcal{H}^\text{eff}$ as a Hamiltonian acting on ion states only,
\begin{equation}
\begin{split}
\label{eq:app_Heff-ions}
&\mathcal{H}^\text{eff}_{MM'} = \langle M | \mathcal{H}^\text{eff} | M' \rangle = \langle M \otimes \Gamma_0 | \mathcal{H}' | \Gamma_0 \otimes M' \rangle 
+ \sum_{\Gamma'' \neq \Gamma_0} \frac{ \langle M \otimes \Gamma_0 | \mathcal{H} | \Gamma''  \rangle \langle \Gamma'' | \mathcal{H} | \Gamma_0 \otimes M' \rangle }{ E_{\Gamma_0} - E_{\Gamma''} } \ .
\end{split}
\end{equation}

Since the $s\pd$ exchange term in $\mathcal{H}$, which produces the non-diagonal matrix element $\langle M \otimes \Gamma_0 | \mathcal{H} | \Gamma'' \otimes M'' \rangle$, is the interaction of a single carrier with an ion, the only $\Gamma''$ states which have a non-zero contribution to the sum over $\Gamma''$ in Eq.~\eqref{eq:app_Heff-ions} are those which are created from $\Gamma_0$ by just one excitation, $\psi_{\vec{k} m} \rightarrow \psi_{\vec{k}'m'}$ from below to above the Fermi level---we denote such states by $\Gamma_{\vec{k}m\vec{k}'m'}$. Hence, we have $E_{\Gamma_0} - E_{\Gamma_{\vec{k}m\vec{k}'m'}} = E_{\vec{k} m} - E_{\vec{k}'m'}$ and Hamiltonian~\eqref{eq:app_Heff-ions} can be written as
\begin{equation}
\label{eq:app_Heff-ions2}
\begin{split}
&\mathcal{H}^\text{eff}_{MM'} = \langle M \otimes \Gamma_0 | \mathcal{H}' | \Gamma_0 \otimes M' \rangle + \sum_{\vec{k}\vec{k}'} \sum_{mm'} \frac{f_{\vec{k} m} (1 - f_{\vec{k}'m'})}{ E_{\vec{k} m} - E_{\vec{k}'m'} }
\langle M \otimes \Gamma_0 | \mathcal{H} | \Gamma_{\vec{k}m\vec{k}'m'}  \rangle \langle \Gamma_{\vec{k}m\vec{k}'m'} | \mathcal{H} | \Gamma_0 \otimes M' \rangle\ ,
\end{split}
\end{equation}
where $f_{\vec{k} m}$ is the Fermi-Dirac distribution (this generalises the calculation to finite temperatures).

To integrate out the carrier degrees of freedom in~\eqref{eq:app_Heff-ions2}, we need to calculate the matrix elements $\langle \psi_{\vec{k}m} | s^\sigma I(\vec{r} - \vec{R}_j) | \psi_{\vec{k}'m'} \rangle$, where $\sigma = +,-,z$ and $[\hat s^+, \hat s^-] = \hat s^z$ by convention. Assuming that the function $I(\vec{r}-\vec{R}_j)$ vanishes quickly outside the unit cell and that $e^{i\vec{k}\cdot\vec{r}}$ and $e^{i\vec{k}'\cdot\vec{r}}$ vary slowly over the same range, we obtain
\[
\langle \psi_{\vec{k}m} | s^\sigma I(\vec{r} - \vec{R}_j) | \psi_{\vec{k}'m'} \rangle = e^{i(\vec{k}'-\vec{k})\cdot\vec{R}_j} \langle u_{\vec{k}m} | s^\sigma I(\vec{r} - \vec{R}_j) | u_{\vec{k}'m'} \rangle \ ,
\]
Because of the spin-orbit coupling we cannot separate $\langle u_{\vec{k}m} | s^\sigma I(\vec{r} - \vec{R}_j) | u_{\vec{k}'m'} \rangle$ into a product of the spin and spatial matrix elements. To overcome this problem, we write $u_{\vec{k}m}$ as $\sum_s u^s_{\vec{k}m} \psi_s$ using the spinor basis $\psi_s = \spu,\spd$, where $u^s_{\vec{k}m}$ is a purely spatial wavefunction, and
\[
\langle u_{\vec{k}m} | s^\sigma I(\vec{r} - \vec{R}_j) | u_{\vec{k}'m'} \rangle = \sum_{ss'} \langle \psi_s | s^\sigma | \psi_{s'} \rangle \langle u^s_{\vec{k}m} | I(\vec{r}-\vec{R}_j) | u^{s'}_{\vec{k}'m'} \rangle \ .
\]
The operator $\Pex$ projects the wavefunctions on the basis functions which are coupled to the magnetic ion via the exchange interaction described by the function $I$. Hence, we can write that
\[
\langle u^s_{\vec{k}m} | I(\vec{r}-\vec{R}_j) | u^{s'}_{\vec{k}'m'} \rangle = c \langle u^s_{\vec{k}m} | \Pex | u^{s'}_{\vec{k}'m'} \rangle \ ,
\]
which leads to
\begin{equation}
\label{eq:Heff-smel1}
\begin{split}
&\langle u_{\vec{k}m} | s^\sigma I(\vec{r} - \vec{R}_j) | u_{\vec{k}'m'} \rangle = c \sum_{ss'} \langle \psi_s | s^\sigma | \psi_{s'} \rangle \langle u^s_{\vec{k}m} | \Pex | u^{s'}_{\vec{k}'m'} \rangle = c \langle u_{\vec{k}m} | s^\sigma \Pex | u_{\vec{k}'m'} \rangle \ ,
\end{split}
\end{equation}
as $\Pex$ commutes with the spin operator. The scaling factor $c$ can be derived from the condition that for the system fully polarised along the $z$ axis and in the mean-field and virtual-crystal approximation, we have
\[
S \sum_{j=1}^N \langle u_{\vec{k}m} | s^z I(\vec{r} - \vec{R}_j) | u_{\vec{k}'m'} \rangle = \Delta \langle u_{\vec{k}m} | s^z \Pex | u_{\vec{k}'m'} \rangle \ ,
\]
which can be simply generalised to arbitrary temperature by replacing $\Delta$ and $S$ by their temperature-dependent versions. Inserting~\eqref{eq:Heff-smel1}, we obtain
\[
NSc\,\langle u_{\vec{k}m} | s^z \Pex | u_{\vec{k}'m'} \rangle = \Delta \langle u_{\vec{k}m} | s^z \Pex | u_{\vec{k}'m'} \rangle \ .
\]
But $\Delta = N S \beta / V$, where $\beta$ is the $\pd$ exchange integral defined in Eq.~\eqref{eq:betadef}, so $c = \Delta / (NS) = \beta / V$ and
\[
\langle \psi_{\vec{k}m} | s^\sigma I(\vec{r} - \vec{R}_j) | \psi_{\vec{k}'m'} \rangle = \frac{\beta}{V} e^{i(\vec{k}'-\vec{k})\cdot\vec{R}_j} s^\sigma_{\vec{k}m\vec{k}'m'} \ .
\]
where $s^\sigma_{\vec{k}m\vec{k}'m'} = \langle u_{\vec{k} m} | s^\sigma \Pex | u_{\vec{k}'m'} \rangle$. The above expression is proportional to $\beta$, but the operator $\Pex$ inside the matrix element takes into account the difference between $\pd$ and $\sd$ exchange integrals.

We can now write Hamiltonian~\eqref{eq:app_Heff-ions} using ion spin operators,
\begin{equation}
\label{eq:app_heisenberg}
\mathcal{H}^\text{eff} = \sum_{\sigma} \sum_{j=1}^{N} H_j^\sigma S_j^{\sigma} + \sum_{\sigma\sigma'} \sum_{j=1}^{N} \sum_{j'=1}^{N} H_{jj'}^{\sigma\sigma'} S_j^{\sigma} S_{j'}^{\sigma'}\ .
\end{equation}
The coefficients $H_j^\sigma$ and $H_{jj'}^{\sigma\sigma'}$ are given by
\begin{equation}
\label{eq:app_hjjss}
H_{jj'}^{\sigma\sigma'} = \frac{\beta^2}{V^2} \sum_{\vec{k}\vec{k'}} \sum_{mm'} \frac{f_{\vec{k} m} (1 - f_{\vec{k}'m'})}{E_{\vec{k}'m'} - E_{\vec{k} m}}
e^{i(\vec{k'}-\vec{k})\cdot(\vec{R}_j - \vec{R}_{j'})} s^{\sigma}_{\vec{k}m\vec{k}'m'} s^{\sigma'}_{\vec{k'}m'\vec{k}m}\ ,
\end{equation}
where due to the condition $\Gamma'' \neq \Gamma_0$ in Eq.~\eqref{eq:app_Heff2}, for $\vec k = \vec k'$ the summation goes over $m \neq m'$, and
\begin{equation}
\label{eq:app_hjs}
\begin{split}
H_j^\sigma &= \frac{\beta}{V} \sum_\vec{k} \sum_m f_{\vec{k} m} s^{\sigma}_{\vec{k}m\vec{k}m} - \frac{\Delta \beta}{V} \sum_\vec{k} \sum_{m \neq m'} \frac{f_{\vec{k} m} (1 - f_{\vec{k},m'})}{E_{\vec{k},m'} - E_{\vec{k} m}} ( s^\sigma_{\vec{k}m\vec{k}m'} s^z_{\vec{k}m'\vec{k}m} + s^\sigma_{\vec{k}m'\vec{k}m} s^z_{\vec{k}m\vec{k}m'} ) \ .
\end{split}
\end{equation}
To obtain the above expressions, we substituted the carrier-only part of Ha\-mil\-ton\-ian $\mathcal{H}_0$~\eqref{eq:app_Hamiltonian} by the sum of $P$ one-par\-ti\-cle Ha\-mil\-ton\-ians $h_0$ from Eq.~\eqref{eq:app_H1part}. We also used the formula $\langle \psi_{\vec{k} m} | h_0 | \psi_{\vec{k}'m'} \rangle = \delta_{\vec{k}\vec{k}'} (\delta_{mm'} E_{\vec{k} m} - \Delta s^z_{\vec{k}m\vec{k}'m'})$, obtained from Hamiltonian $h$~\eqref{eq:app_H1part}.

Using the L\"owdin perturbation-variational calculus, we have thus described the problem as a lattice spin system coupled by the effective exchange interaction, integrating out the carrier degrees of freedom. The physics of the carriers is embedded in the effective Hamiltonian $\mathcal{H}^\text{eff}$~\eqref{eq:app_heisenberg}, and is responsible for the long-range nonlocal character of the mutual interactions between magnetic ions.

The first term of $\mathcal{H}^\text{eff}$ contains the operator responsible for the mean-field generated by the carriers acting on the lattice ions (the first part in~\ref{eq:app_hjs}), plus the correction arising from the inter-band transitions. The latter can be associated with the Bloembergen--Rowland exchange mechanism.~\cite{bloembergen, Kacman2001} A small but worth noting result of the next section is that in the small oscillations approximation it cancels with the contribution from the second term.

The second term of $\mathcal{H}^\text{eff}$ describes the effective long-range exchange interaction between the lattice ions. The nature of the interaction, mediated by the carrier between two ions, is reflected in the appearance of the squared $\beta$ constant. The $\sd$ exchange interaction characterised by the $\alpha$ constant is incorporated in the $s^{\sigma}_{\vec{k}m\vec{k}'m'}$ matrix elements via the $\Pex$ operator. The fraction with the resonance denominator resulting from the perturbational approach dampens the influence of the distant energy bands, while the biggest contribution to the sum comes from the states in the vicinity of the Fermi level $E_F$. The denominator looks dangerous, as it may cause the fraction to diverge in the presence of the energy bands' crossings, which would make our perturbation calculus invalid. However, the effective Hamiltonian for ions depends on the average of these factors, and will be shown (at least in the small oscillations approximation) immune to this problem in Sec.~\ref{sec:smallosc}.

Contrary to the original RKKY range function
\[
J (r) = -\frac{\rho(E_F) k_\text{F}^3 J_0^2}{2\pi} \frac{\sin(2 k_\text{F} r) - 2 k_\text{F} r \cos(2 k_\text{F} r)}{(2 k_\text{F} r)^4}
\]
(where $\rho(E_F)$ and $k_\text{F}$ are the density of states at the Fermi level and the Fermi wavevector of the carriers), the formula~\eqref{eq:app_hjjss} is anisotropic in space in the presence of the spin-orbit coupling, reflecting the symmetries of the crystal lattice. If, additionally, the bulk and structure inversion symmetries are broken, it has an antisymmetric part in the form of the Dzyaloshinskii--Moriya interaction~\cite{Dzyaloshinskii:1958_JPCS, Moriya:1960_PR},
\begin{equation}
	\label{eq:app_DMH}
	\sum_{\sigma^{\bis}} \sum_{jj'} i u_{jj'}^{\sigma^{\bis}} \sum_{\sigma\sigma'} \epsilon_{\sigma^{\bis}\sigma\sigma'} S_j^\sigma S_{j'}^{\sigma'}\ ,
\end{equation}
where $\epsilon_{\sigma^{\bis}\sigma\sigma'}$ is the antisymmetric Levi-Civita symbol with  $\epsilon_{{\psms} z} = 1$, while $\vec u_{ij}$ is a pseudovector and exists only in the systems with broken inversion symmetry,
\[
i u_{jj'}^{\sigma^\bis} = \frac{1}{2} \sum_{\sigma\sigma'} \epsilon_{{\sigma^\bis}\sigma\sigma'} \
H_{jj'}^{\sigma\sigma'}\ .
\]

\section{Small oscillations approximation}
\label{sec:smallosc}

The system of coupled magnetic moments acts like that of harmonic oscillators, an analogy which is concretised mathematically by the Holstein-Primakoff bosonisation~\cite{Holstein:1940_PR}:
\[
\begin{split}
S_j^+ \approx \sqrt{S} \sqrt{ 1 - \frac{a_j^\dagger a_j}{2S} } a_j \ ,\ 
S_j^- \approx \sqrt{S} a_j^\dagger \sqrt{ 1 - \frac{a_j^\dagger a_j}{2S} } \ ,\ 
S_j^z = S - a_j^\dagger a_j\ ,
\end{split}
\]
which replaces the spin operators with nonlinear functions of bosonic creation and annihilation operators $a_j^\dagger$ and $a_j$. To investigate the dynamics of the groundstate and low-lying excitations of the effective Hamiltonian~\eqref{eq:app_heisenberg}, we will use the small oscillations approximation and approximate these functions with their power expansions around the state of saturation magnetisation:
\begin{equation}
\label{eq:app_spinoperators}
S_j^+ \approx \sqrt{S} a_j \ ,\ \ S_j^- \approx \sqrt{S} a_j^\dagger \ ,\ \ S_j^z = S - a_j^\dagger a_j\ ,
\end{equation}
leaving in the Hamiltonian only those terms which are quadratic in the creation and annihilation operators.
\begin{equation}
\begin{split}
\label{eq:app_Heff-ions-quadr}
&\mathcal{H}^\text{eff} =  - \frac{\beta}{V}\ \sum_{j=1}^{N} \sum_\vec{k}  \left( \sum_m f_{\vec{k} m} s^z_{\vec{k}m\vec{k}m} - 2 \Delta  \sum_{m \neq m'} \frac{f_{\vec{k} m} (1 - f_{\vec{k},m'})}{E_{\vec{k},m'} - E_{\vec{k} m}} 
|s^z_{\vec{k}m\vec{k}m'}|^2  \right) a_j^\dagger a_j \\
&- \frac{S \beta^2}{V^2} \sum_{j,j'=1}^{N} \left( \, \sum_{\vec{k}\vec{k'}mm'} \frac{f_{\vec{k} m} (1 - f_{\vec{k}'m'})}{E_{\vec{k}'m'} - E_{\vec{k} m}}  e^{i(\vec{k'}-\vec{k})\cdot(\vec{R}_j - \vec{R}_{j'})} |s^z_{\vec{k}m\vec{k}'m'}|^2 \right)( a_j^\dagger a_j + a_{j'}^\dagger a_{j'} ) \\
&+ \frac{S \beta^2}{V^2} \sum_{j=1}^{N} \sum_{j'=1}^{N}  \sum_{\vec{k}\vec{k'}} \sum_{mm'} \frac{f_{\vec{k} m} (1 - f_{\vec{k}'m'})}{E_{\vec{k}'m'} - E_{\vec{k} m}}  e^{i(\vec{k'}-\vec{k})\cdot(\vec{R}_j - \vec{R}_{j'})} ( s^+_{\vec{k}m\vec{k}'m'} a_j + s^-_{\vec{k}m\vec{k}'m'} a^\dagger_j )^\dagger ( s^+_{\vec{k}m\vec{k}'m'} a_j + s^-_{\vec{k}m\vec{k}'m'} a^\dagger_j ) \ .
\end{split}
\end{equation}
Within the small oscillations approximation, the linear and zeroth-order terms do not affect the excitation spectrum. This approximation works very well in the long-wave limit, $a q \ll \pi$, as the neglected magnon-magnon interactions are proportional to $(a q)^4$~\cite{Kittel:1987_B}. 

Firstly, let us take a closer look at the term
\[
\frac{S \beta^2}{V^2} \sum_{j,j'=1}^{N} \left( \sum_{\vec{k}\vec{k'}} \sum_{mm'} \frac{f_{\vec{k} m} (1 - f_{\vec{k}'m'})}{E_{\vec{k}'m'} - E_{\vec{k} m}}  e^{i(\vec{k'}-\vec{k})\cdot(\vec{R}_j - \vec{R}_{j'})} |s^z_{\vec{k}m\vec{k}'m'}|^2 \right)
a_j^\dagger a_j \ .
\]
Summation over $j'$ gives $\sum_{j'=1}^N e^{i(\vec{k'}-\vec{k})\cdot\vec{R}_{j'}} = N \delta_{\vec{k}\vec{k}'}$, which allows us to write the above term as
\[
\frac{N S \beta^2}{V^2} \sum_{j=1}^{N} \left( \sum_{\vec{k}} \sum_{mm'} \frac{f_{\vec{k} m} (1 - f_{\vec{k},m'})}{E_{\vec{k},m'} - E_{\vec{k} m}}  |s^z_{\vec{k}m\vec{k}m'}|^2 \right)
a_j^\dagger a_j \ .
\]
The twin term with $a_{j'}^\dagger a_{j'}$ at the end is treated in the same fashion, after which we write~\eqref{eq:app_Heff-ions-quadr} as
\[
\begin{split}
&\mathcal{H}^\text{eff} = - \frac{\beta}{V} \sum_{j=1}^{N} \sum_\vec{k} \left( \sum_m f_{\vec{k} m} s^z_{\vec{k}m\vec{k}m} - 2 \Delta  \sum_{m \neq m'} \frac{f_{\vec{k} m} (1 - f_{\vec{k},m'})}{E_{\vec{k},m'} - E_{\vec{k} m}} 
|s^z_{\vec{k}m\vec{k}m'}|^2  \right) a_j^\dagger a_j \\
&- \frac{2 N S \beta^2}{V^2} \sum_{j=1}^{N} \left( \sum_{\vec{k}} \sum_{mm'} \frac{f_{\vec{k} m} (1 - f_{\vec{k},m'})}{E_{\vec{k},m'} - E_{\vec{k} m}}  |s^z_{\vec{k}m\vec{k}m'}|^2 \right)
a_j^\dagger a_j \\
&+ \frac{S \beta^2}{V^2} \sum_{j=1}^{N} \sum_{j'=1}^{N}  \sum_{\vec{k}\vec{k'}} \sum_{mm'} \frac{f_{\vec{k} m} (1 - f_{\vec{k}'m'})}{E_{\vec{k}'m'} - E_{\vec{k} m}}  e^{i(\vec{k'}-\vec{k})\cdot(\vec{R}_j - \vec{R}_{j'})} ( s^+_{\vec{k}m\vec{k}'m'} a_j + s^-_{\vec{k}m\vec{k}'m'} a^\dagger_j )^\dagger ( s^+_{\vec{k}m\vec{k}'m'} a_j + s^-_{\vec{k}m\vec{k}'m'} a^\dagger_j ) \ .
\end{split}
\]
Due to the equality $\Delta = N S \beta / V$, terms with $|s^z_{\vec{k}m\vec{k}m'}|^2$ cancel and we get
\[
\begin{split}
&\mathcal{H}^\text{eff} =  - \frac{\beta}{V} \left( \sum_\vec{k} \sum_m f_{\vec{k} m} s^z_{\vec{k}m\vec{k}m} \right) \sum_{j=1}^{N} a_j^\dagger a_j  + \\
& \frac{S \beta^2}{V^2}    \sum_{\vec{k}\vec{k'}} \sum_{mm'} \frac{f_{\vec{k} m} (1 - f_{\vec{k}'m'})}{E_{\vec{k}'m'} - E_{\vec{k} m}} \left[ \sum_{j=1}^{N} e^{i(\vec{k'}-\vec{k})\cdot\vec{R}_j} ( s^+_{\vec{k}m\vec{k}'m'} a_{j} + s^-_{\vec{k}m\vec{k}'m'} a^\dagger_j ) \right] \left[ \sum_{j'=1}^{N} e^{i(\vec{k'}-\vec{k})\cdot\vec{R}_{j'}} ( s^+_{\vec{k}m\vec{k}'m'} a_{j'} + s^-_{\vec{k}m\vec{k}'m'} a^\dagger_{j'} ) \right]^\dagger .
\end{split}
\]

We can now proceed with the Fourier transform (invoking the virtual-crystal approximation), using the fact that $\sum_{j=1}^{N} a_j^\dagger a_j = \sum_{\vec{q}} a_{\vec{q}}^\dagger a_{\vec{q}}$:
\[
\begin{split}
&\mathcal{H}^\text{eff} =  - \frac{\beta}{V} \left( \sum_\vec{k} \sum_m f_{\vec{k} m} s^z_{\vec{k}m\vec{k}m} \right) \sum_{\vec{q}} a_{\vec{q}}^\dagger a_{\vec{q}} \\ 
&+ \frac{\Delta \beta}{V}  \sum_{\vec{k}\vec{k'}} \sum_{mm'} \frac{f_{\vec{k} m} (1 - f_{\vec{k}'m'})}{E_{\vec{k}'m'} - E_{\vec{k} m}} \left( s^+_{\vec{k}m\vec{k}'m'} a_{\vec{k'}-\vec{k}} + s^-_{\vec{k}m\vec{k}'m'} a^\dagger_{\vec{k'}-\vec{k}} \right)
 \left( s^+_{\vec{k}m\vec{k}'m'} a_{\vec{k'}-\vec{k}} + s^-_{\vec{k}m\vec{k}'m'} a^\dagger_{\vec{k'}-\vec{k}}  \right)^\dagger \ .
\end{split}
\]

After simple algebraic transformations, including the symmetrisation of the sums over wavevectors, we arrive at the final form of the harmonic Hamiltonian,
\begin{equation}
\label{eq:app_Ha}
\mathcal H^\text{eff} = \sum_{\vec{q}} \Bigl[ \left( \Xi - \chi_{\vec{q}}^{\psms} \right) a_{\vec{q}}^\dagger a_{\vec{q}} - \frac{1}{2} \chi_{\vec{q}}^{\psps} a_{\vec{q}} a_{-\vec{q}} - \frac{1}{2} \chi_{\vec{q}}^{\msms} a_{\vec{q}}^\dagger a_{-\vec{q}}^\dagger \Bigr] \ .
\end{equation}
We call it the \textit{interaction} representation as it describes the perturbation of the ground state by the isotropic Coulomb interaction (first term) and by the spin-orbit interaction, coupling modes of different $\vec{q}$ (remaining terms). The spin susceptibility of the carriers is given by
\begin{equation}
\begin{split}
\label{eq:app_chi}
\chi^{\sigma\sigma'}_\vec{q} = &-\frac{n S \beta^2}{V} \sum_{\vec{k}} \sum_{mm'} \frac{f_{\vec{k} m} - f_{\vec{k}+\vec{q},m'}}{E_{\vec{k} m} - E_{\vec{k}+\vec{q},m'}}  s^{\sigma}_{\vec{k}m(\vec{k}+\vec{q})m'} s^{\sigma'}_{(\vec{k}+\vec{q})m'\vec{k}m}\ ,
\end{split}
\end{equation}
where $n=N/V$ is the density of localised spins $S$ in the sample volume $V$ and $n S \beta = \Delta$. The presence of the energy denominator shows that $\chi^{\sigma\sigma'}_\vec{q}$ corresponds to the second-order part of the Hamiltonian~\eqref{eq:app_Heff-ions2}. As promised in Sec.~\ref{sec:selfconloewdin}, the vanishing of the denominator is not harmful, due to the de l'Hospital rule.

The formula~\eqref{eq:app_chi} implies that $\chi^{\psps}_\vec{q} = (\chi^{\msms}_\vec{q})^\ast$ is symmetric in $\vec q$. In the absence of the spin-orbit coupling the bands' spins become fully polarised, which causes $\chi^{\psps}_\vec{q}$ to vanish, as $\langle \psi | s^+ | \psi' \rangle \langle \psi' | s^+ | \psi \rangle = 0$ for any choice of spinors $\psi,\psi' = \spu,\spd$. This is also true for non-zero spin-orbit coupling in the case when the valence bands are isotropic~\cite{Konig:2001_PRB} and for $s$-type bands in general, as for them the total angular momentum is equal to spin and thus they are fully polarised even in the presence of spin-orbit interaction.

Because $\chi^{\psms}_\vec{q}$, $\chi^{\msps}_\vec{q} \in \mathcal R$ inherit the symmetry of the $\psi_{\vec{k} m}$ eigenstates, it can be expected to be symmetric with respect to $\vec q$ for systems which preserve space inversion symmetry, like in the case analysed in Ref.~\onlinecite{Konig:2001_PRB}, and otherwise for systems which do not.
The $\vec{q}$-independent term describes the interaction of a single magnetic ion with a molecular field arising from the intraband spin polarisation of the carriers,
\begin{equation}
\label{eq:app_Xi}
\Xi = -\frac{\beta}{V} \sum_{\vec k} \sum_m f_{\vec k,m}\, s^z_{\vec km\vec km}\ .
\end{equation}
The corresponding term reflecting the interband polarisation,
\begin{equation}
\label{eq:app_Xiso}
\Xiso = \frac{n S \beta^2}{V} \sum_{\vec{k}} \sum_{m\neq m'} \frac{f_{\vec{k} m} - f_{\vec{k},m'}}{E_{\vec{k},m'} - E_{\vec{k} m}} |s^z_{\vec{k}m\vec{k}m'}|^2\ ,
\end{equation}
arises from both $H_{jj'}^{\sigma\sigma'}$~\eqref{eq:app_hjjss} and the second part of $H_j^\sigma$ coefficient~\eqref{eq:app_hjs}, and cancels exactly in the full Hamiltonian $\mathcal H^\text{eff}$.
As announced in the previous section, this shows that the Bloembergen--Rowland exchange, driven by virtual spin transitions between different bands, does not exist in the low-temperature limit covered by small oscillations approximation.

\section{Bogoliubov transform}

So far, we have used the L\"{o}wdin calculus to find the effective Hamiltonian of lattice ions interacting through delocalised carriers in a self-consistent way, written in the interaction representation of creation and annihilation operators. To calculate the dispersion dependence of its low lying energy states, spin waves, we need to transform the Hamiltonian to the \textit{spin-wave representation}. For this purpose we will use the Bogoliubov transform, which we have adapted to systems with inversion symmetry breaking. 

The effective Hamiltonian for the lattice ions derived in Secs.~\ref{sec:selfconloewdin} and~\ref{sec:smallosc} in the interaction picture has the following form:
\[
\mathcal H^\text{eff} = \sum_{\vec{q}} \Bigl[ \left( \Xi - \chi_{\vec{q}}^{\psms} \right) a_{\vec{q}}^\dagger a_{\vec{q}} - \frac{1}{2} \chi_{\vec{q}}^{\psps} a_{\vec{q}} a_{-\vec{q}} - \frac{1}{2} \chi_{\vec{q}}^{\msms} a_{\vec{q}}^\dagger a_{-\vec{q}}^\dagger \Bigr] \ .
\]
It describes the spin system in terms of circularly polarised plane waves (the first term), which interact with each other and deform in time (the remaining terms).

We want to obtain the dispersion relation of independent, stable magnons. For this purpose, we diagonalise $\mathcal H^\text{eff}$ by the Bogoliubov transformation from $a_\vec{q}, a_\vec{q}^\dagger$ to $b_\vec{q}, b_\vec{q}^\dagger$ operators (which describe independent excitation modes), keeping in mind that we deal with the system which breaks the space inversion symmetry. Because $\mathcal H^\text{eff}$ in the above form mixes states with opposite wavevectors, we write the sum over $\vec{q}$ in an explicitly symmetrised form:
\begin{equation}
\label{eq:app_Ha2}
\begin{split}
\mathcal H^\text{eff} &= \frac{1}{2} \sum_{\vec{q}} \Bigl[ 
\left( \Xi - \chi_{\vec{q}}^{\psms} \right) a_{\vec{q}}^\dagger a_{\vec{q}}
+ \left( \Xi - \chi_{-\vec{q}}^{\psms} \right) a_{-\vec{q}}^\dagger a_{-\vec{q}} - \frac{1}{2} (\chi_{\vec{q}}^{\psps} + \chi_{-\vec{q}}^{\psps}) a_{\vec{q}} a_{-\vec{q}} 
- \frac{1}{2} (\chi_{\vec{q}}^{\msms} + \chi_{-\vec{q}}^{\msms}) a_{\vec{q}}^\dagger a_{-\vec{q}}^\dagger 
\Bigr] \ .
\end{split}
\end{equation}
The Bogoliubov transformation is given by the formula
\begin{equation}
\label{eq:app_bogo}
a_\vec{q} = u_\vec{q} b_\vec{q} + v_\vec{q} b_{-\vec{q}}^\dagger \ , \quad u_\vec{q},v_\vec{q} \in \mathbb{C} \ ,
\end{equation}
with the conditions
\begin{equation}
\label{eq:app_uvcond}
\begin{split}
1 = |u_\vec{q}|^2 - |v_\vec{q}|^2 \ , \ \ 0 = u_\vec{q} v_{-\vec{q}} - v_\vec{q} u_{-\vec{q}}
\end{split}
\end{equation}
ensuring the preservation of canonical commutation relations (equivalently, the invertibility of this transformation). Inserting this into~\eqref{eq:app_Ha2}, we obtain (neglecting the lower-order terms and making use of the properties of $\chi$'s)
\begin{equation}
\label{eq:app_Hb}
\begin{split}
&\mathcal H^\text{eff} = \frac{1}{2} \sum_{\vec{q}} \Bigl[ 
\left( \Xi - \chi_{\vec{q}}^{\psms} \right) (|u_\vec{q}|^2 b_\vec{q}^\dagger b_\vec{q} + |v_\vec{q}|^2 b_{-\vec{q}}^\dagger b_{-\vec{q}} + u^\ast_\vec{q} v_\vec{q} b_\vec{q}^\dagger b_{-\vec{q}}^\dagger + u_\vec{q} v_\vec{q}^\ast b_\vec{q} b_{-\vec{q}} ) \\
&\ + \left( \Xi - \chi_{-\vec{q}}^{\psms} \right) (|u_{-\vec{q}}|^2 b_{-\vec{q}}^\dagger b_{-\vec{q}} + |v_{-\vec{q}}|^2 b_{\vec{q}}^\dagger b_{\vec{q}} + u^\ast_{-\vec{q}} v_{-\vec{q}} b_{\vec{q}}^\dagger b_{-\vec{q}}^\dagger + u_{-\vec{q}} v_{-\vec{q}}^\ast b_{\vec{q}} b_{-\vec{q}} ) \\
&\ - \chi_{\vec{q}}^{\psps} ( u_\vec{q} u_{-\vec{q}} b_\vec{q} b_{-\vec{q}} + u_\vec{q} v_{-\vec{q}} b_\vec{q}^\dagger b_\vec{q} + v_\vec{q} u_{-\vec{q}} b_{-\vec{q}}^\dagger b_{-\vec{q}} + v_\vec{q} v_{-\vec{q}} b_\vec{q}^\dagger b_{-\vec{q}}^\dagger )
\\
&\ - (\chi_{\vec{q}}^{\psps})^\ast
( u_\vec{q}^\ast u_{-\vec{q}}^\ast b_\vec{q}^\dagger b_{-\vec{q}}^\dagger + u_\vec{q}^\ast v_{-\vec{q}}^\ast b_\vec{q}^\dagger b_\vec{q} + v_\vec{q}^\ast u_{-\vec{q}}^\ast b_{-\vec{q}}^\dagger b_{-\vec{q}} + v_\vec{q}^\ast v_{-\vec{q}}^\ast b_\vec{q} b_{-\vec{q}} )
\Bigr] \ ,
\end{split}
\end{equation}
which leads to the following complex equation for each $\vec{q}$:
\begin{equation}
\label{eq:app_uveqs}
\left( \Xi - \chi_{\vec{q}}^{\psms} \right) u_\vec{q} v_\vec{q}^\ast  
+  \left( \Xi - \chi_{-\vec{q}}^{\psms} \right) u_{-\vec{q}} v_{-\vec{q}}^\ast 
-  \chi_{\vec{q}}^{\psps} u_\vec{q} u_{-\vec{q}} 
 - (\chi_{\vec{q}}^{\psps})^\ast v_\vec{q}^\ast v_{-\vec{q}}^\ast = 0 \ .
\end{equation}
Since all equations for $u$'s and $v$'s are invariant under the reflection of $\vec{q}$, we assume that $u_\vec{q} = u_{-\vec{q}}$ and $v_\vec{q} = v_{-\vec{q}}$. The standard parametrisation for $u_\vec{q},v_\vec{q}$ consistent with the first condition~\eqref{eq:app_uvcond} reads
\[
u_\vec{q} = e^{i \mu_\vec{q}} \cosh \theta_\vec{q} \ ,\ \ v_\vec{q} = e^{i \nu_\vec{q}} \sinh \theta_\vec{q} \ .
\]
Inserting it into~\eqref{eq:app_uveqs} gives
\[
\begin{split}
0 &= \left( 2 \Xi - \chi_{\vec{q}}^{\psms} - \chi_{-\vec{q}}^{\psms} \right) \cosh\theta_\vec{q} \sinh\theta_\vec{q} e^{i(\mu_\vec{q} - \nu_\vec{q})} - \chi_{\vec{q}}^{\psps} \cosh^2\theta_\vec{q} e^{2 i \mu_\vec{q}} - (\chi_{\vec{q}}^{\psps})^\ast  \sinh^2\theta_\vec{q} e^{-2i\nu_\vec{q}} \\
&= \left( 2 \Xi - \chi_{\vec{q}}^{\psms} - \chi_{-\vec{q}}^{\psms} \right) \cosh\theta_\vec{q} \sinh\theta_\vec{q} - \chi_{\vec{q}}^{\psps} \cosh^2\theta_\vec{q} e^{i(\mu_\vec{q} + \nu_\vec{q})} - (\chi_{\vec{q}}^{\psps})^\ast  \sinh^2\theta_\vec{q} e^{-i(\mu_\vec{q} + \nu_\vec{q})} \ .
\end{split}
\]
There is no equation for $\mu_\vec{q} - \nu_\vec{q}$, so we assume it is zero. We can replace $\mu_\vec{q}$ by another unknown, $\mu'_\vec{q}$, defined by
\[
\chi_{\vec{q}}^{\psps} e^{2 i \mu_\vec{q}} = |\chi_{\vec{q}}^{\psps}| e^{2 i \mu'_\vec{q}} \ ,
\]
resulting in
\[
\left( 2 \Xi - \chi_{\vec{q}}^{\psms} - \chi_{-\vec{q}}^{\psms} \right) \cosh\theta_\vec{q} \sinh\theta_\vec{q} = | \chi_{\vec{q}}^{\psps} | \left( \cosh^2\theta_\vec{q} e^{2i\mu'_\vec{q}} + \sinh^2\theta_\vec{q} e^{-2i\mu'_\vec{q}} \right)
\]
or
\[
\begin{split}
&\left( 2 \Xi - \chi_{\vec{q}}^{\psms} - \chi_{-\vec{q}}^{\psms} \right) \cosh\theta_\vec{q} \sinh\theta_\vec{q} = | \chi_{\vec{q}}^{\psps} | \left( \cosh^2\theta_\vec{q} + \sinh^2\theta_\vec{q} \right) \cos 2\mu'_\vec{q} + i \sin 2\mu'_\vec{q} \ ,
\end{split}
\]
which requires $\mu'_\vec{q} = 0$, leading to the equation
\[
\left( 2 \Xi - \chi_{\vec{q}}^{\psms} - \chi_{-\vec{q}}^{\psms} \right) \cosh\theta_\vec{q} \sinh\theta_\vec{q} = |\chi_{\vec{q}}^{\psps}| ( \cosh^2\theta_\vec{q} + \sinh^2\theta_\vec{q} ) \ ,
\]
hence
\begin{equation}
\label{eq:app_thetaq}
2 \theta_\vec{q} = \arctanh \frac{|\chi_{\vec{q}}^{\psps}|}{ \Xi - \tfrac{1}{2} (\chi_{\vec{q}}^{\psms} + \chi_{-\vec{q}}^{\psms})} \ ,
\end{equation}
with the solution non-existent when $\left| \frac{|\chi_{\vec{q}}^{\psps}|}{ \Xi - \tfrac{1}{2} (\chi_{\vec{q}}^{\psms} + \chi_{-\vec{q}}^{\psms})} \right| > 1$. This particular\footnote{Other solutions may exist which do not fulfil our assumption that $u_\vec{q} = u_{-\vec{q}}$ and $v_\vec{q} = v_{-\vec{q}}$. However, for a finite number of degrees of freedom the representation of canonical commutation relations is unique up to a unitary transformation, which guarantees the uniqueness of the obtained dispersion relation. In general, a set of creation and annihilation operators diagonalising $\mathcal H^\text{eff}$ must exist, but it does not need to be obtainable via a transformation of the form~\eqref{eq:app_bogo}.} solution will turn out to be sufficient for our needs, as numerical calculations give $|\chi_{\vec{q}}^{\psps}| \ll | \Xi - \tfrac{1}{2} (\chi_{\vec{q}}^{\psms} + \chi_{-\vec{q}}^{\psms}) |$.

Gathering all terms multiplying $b_\vec{q}^\dagger b_\vec{q}$ in~\eqref{eq:app_Hb}, we obtain
\[
\begin{split}
& \frac{1}{2} \Bigl[ \left( \Xi - \chi_{\vec{q}}^{\psms} \right) |u_\vec{q}|^2 
+ \left( \Xi - \chi_{-\vec{q}}^{\psms} \right) |v_{-\vec{q}}|^2 
- \chi_{\vec{q}}^{\psps}  u_\vec{q} v_{-\vec{q}} - (\chi_{\vec{q}}^{\psps})^\ast u_\vec{q}^\ast v_{-\vec{q}}^\ast \Bigr] \\
&= \frac{1}{2} \Bigl[ \left( \Xi - \chi_{\vec{q}}^{\psms} \right) \cosh^2\theta_\vec{q}
+ \left( \Xi - \chi_{-\vec{q}}^{\psms} \right) \sinh^2\theta_{\vec{q}} - 2 |\chi_{\vec{q}}^{\psps}| \cosh \theta_\vec{q} \sinh \theta_{\vec{q}} \Bigr] \\
&= \frac{1}{2} \Bigl[ \left( \Xi - \frac{ \chi_{\vec{q}}^{\psms} + \chi_{-\vec{q}}^{\psms} }{2} \right) \cosh 2\theta_\vec{q} - \frac{ \chi_{\vec{q}}^{\psms} - \chi_{-\vec{q}}^{\psms} }{2} - |\chi_{\vec{q}}^{\psps}| \sinh 2\theta_\vec{q} \Bigr] \ .
\end{split}
\]
Using~\eqref{eq:app_thetaq} and assuming that $ \Xi - \frac{ \chi_{\vec{q}}^{\psms} + \chi_{-\vec{q}}^{\psms} }{2} > 0$ (which is the case for our numerical calculations), we obtain the final form of the effective Hamiltonian in the \textit{spin-wave representation}:
\begin{equation}
\label{eq:app_Hbb}
\mathcal{H}^\text{eff} = -\sum_{\vec{q}} \omega_{\vec{q}} b_\vec{q}^{\dagger} b_\vec{q} \ ,
\end{equation}
where excitation modes are spin waves with dispersion\footnote{The negative sign of the Hamiltonian~\eqref{eq:app_Hbb} is a consequence of using the electronic convention to describe the carrier-ion system.}
\begin{equation}
\label{eq:app_omega}
\omega_\vec{q} = \frac{\chi_{\text{\small{-}}\vec{q}}^{\psms} {\scriptstyle -} \chi_\vec{q}^{\psms}}{2} + \sqrt{ \frac{(2\Xi {\scriptstyle -} \chi_\vec{q}^{\psms} {\scriptstyle -} \chi_{\text{\small{-}}\vec{q}}^{\psms})^2}{4} - |\chi_\vec{q}^{\psps}|^2}\,.
\end{equation}
In the case of $\chi_\vec{q}^{\psms} = \chi_{-\vec{q}}^{\psms}$, fulfilled for the systems invariant under space inversion, the above formula simplifies to the solution by K\"{o}nig~\textit{et al.}~\cite{Konig:2001_PRB},
\[
\omega_{\vec{q}} = \sqrt{\left(\Xi - \chi_{\vec{q}}^{\psms}\right)^2 - |\chi_{\vec{q}}^{\psps}|^2}\ .
\]
Furthermore, neglecting the spin-orbit coupling, when $\chi^{\psps} = 0$, Bogoliubov transformation is unnecessary, and the effective Hamiltonian is already diagonalised by $a_\vec{q}$, $a_\vec{q}^\dagger$ operators. 


\end{document}